\documentclass{emulateapj}
\usepackage{natbib}
\usepackage{threeparttable}
\usepackage{array,multirow}
\usepackage{booktabs}
\usepackage{bigstrut}
\shorttitle{Improved method to determine the integrated properties of nuclear rings}
\shortauthors{Chao Ma, Richard de Grijs, and Luis C. Ho}
\slugcomment{Accepted for publication in The Astrophysical Journal Supplements}
\begin{document}
\title{IMPROVED METHOD TO DETERMINE THE INTEGRATED PROPERTIES OF NUCLEAR RINGS: NGC 1512}

\author{Chao Ma\altaffilmark{1}, Richard de Grijs\altaffilmark{1,2},
  and Luis C. Ho\altaffilmark{1}}
\altaffiltext{1}{Kavli Institute for Astronomy \& Astrophysics and
  Department of Astronomy, Peking University, Yi He Yuan Lu 5, Hai
  Dian District, Beijing 100871, China; machao@pku.edu.cn}
\altaffiltext{2}{International Space Science Institute--Beijing, 1
  Nanertiao, Zhongguancun, Hai Dian District, Beijing 100190, China}

\begin{abstract}
The integrated properties of nuclear rings are correlated with their
host galaxy's secular evolution and its dynamics, as well as with the
formation and evolution of the ring's star cluster population(s). Here
we present a new method to accurately measure the spectral energy
distribution and current star-formation rate (SFR) of the nuclear ring
in the barred spiral galaxy NGC 1512 based on high-resolution {\sl
  Hubble} and {\sl Spitzer Space Telescope} images. Image degradation
does not have a significant negative effect on the robustness of the
results. To obtain the ring's SFR for the period spanning $\sim$3--10
Myr, we apply our method to the continuum-subtracted H$\alpha$ and 8
$\mu$m images. The resulting SFR surface density, $\Sigma_{\rm
  SFR}$=$0.09\, {M}_{\odot} \,{\rm yr}^{-1}$ ${\rm kpc}^{-2}$, which
is much higher than the disk-averaged SFR densities in normal
galaxies. We also estimate the ring's total stellar mass, log
(${M}/{M}_{\odot}$) = 7.1 $\pm$ 0.11 for an average age of $\sim$40
Myr.
\end{abstract}

\section{INTRODUCTION}

Circumnuclear starburst rings, which are usually found within 1 kpc of
galactic nuclei, are interesting substructures that are often subject
to intense starbursts. They are mostly found in barred spiral
galaxies, since they are the natural products of bar-induced dynamics,
although it has been reported that merger events and other
non-axisymmetric structures---such as oval/elliptically shaped density
enhancements and strong spiral arms---in some unbarred galaxies can
also generate nuclear rings \citep[e.g., NGC 7217 and NGC
  7742;][]{shlosman1990,Athan1994,combes2001,kormendy2004,mazzuca2006,silchenko2006}. The
gravitational torque from the asymmetric bar potential causes gas
infall toward the nucleus along the dust lanes at the leading edges of
the bar \citep[e.g.,][]{Byrd1994,Knapen1995,Regan2003,Kim2012a}. 
  The inflowing materials, which lose most of their angular momentum,
spiral in toward the ring region at the two `contact points' between
the dust lanes and the ring, where the accumulated gas proceeds to
move in nearly circular orbits and forms a luminous, compact ring
around the galactic center
\citep[e.g.,][]{Athan1992,Buta1996,Mac2004,Kim2012b}.

The high gas density
\citep[e.g.,][]{wild1992,mazzuca2008,comeron2010}, the unusually short
crossing timescale, and the enhanced collision rate in the ring may
consequently trigger violent starburst episodes, which often dominate
the entire star-formation history of the host galaxy
\citep[e.g.,][]{genzel1995,Knapen2006,mazzuca2011}. Comprehensive
studies of nuclear rings are therefore paramount to understand the
secular evolution and dynamics of spiral galaxies. Over the course of
several decades, many theoretical and observational studies have
contributed significantly to our knowledge of nuclear ring properties
\citep[e.g.,][]{burbidge1960,sandage1961,hummel1987,buta1993,englmaier2000,van2013}. Owing
to the close connection between rapid star formation and the presence
of young star clusters, nuclear rings are prime regions in nearby disk
galaxies where young massive star clusters are found in abundance
\citep[e.g.,][]{barth1995,maoz1996,hsieh2012,grijs2012,van2013,grijs2017}.

The properties of nuclear rings are also good diagnostics to constrain
the host galaxy's structural parameters
\citep[e.g.,][]{weiner2001,kormendy2004,Li2015}. They thus serve as
important features connecting galaxy-scale properties to the star
cluster population. Since our ultimate purpose is to explore how the
ring environment (or the integrated ring properties) affects the
evolution of the young cluster population, it is of great importance
to derive the integrated ring parameters, such as their total
luminosities and their star-formation rates (SFRs). Traditional
approaches to calculate the global ring SFR
\citep[e.g.,][]{buta1999,maoz2001,mazzuca2008} usually follow
\citet[][]{kennicutt1998}:
\begin{equation}
\rm{SFR}\, (M_{\odot}\, yr^{-1}) = 7.9 \times 10^{-42} [{\it L} (H\alpha)_{\rm{obs}}],
\end{equation}
where $L$(H$\alpha$)$_{\rm obs}$ represents the H$\alpha$
emission-line luminosity, which is usually directly derived from the
imaging. However, as readily recognized by these authors,
  application of this equation could result in large uncertainties of
  up to 50\% \citep[e.g.,][]{kennicutt1998} because of the presence of
  variable extinction across the ring, although some investigators
have tried to minimize this effect by adopting a representative ring
cluster extinction value \citep[e.g.,][]{buta2000,benedict2002}. To
circumvent the dust-attenuation problem, it is indispensable to
combine observations of ultraviolet or optical SFR tracers with
complementary infrared measurements so as to account for missing flux
caused by the dust absorption and scattering \citep[for a review, see
  e.g.][and references therein]{kennicutt2012}.

\citet[][]{mazzuca2008} have shown that nuclear rings are coplanar
with their disks and retain nearly circular shapes after
deprojection. Since these structures are embedded in the galactic disk
and bulge of their host galaxy, the total, intrinsic ring luminosities
measured are unavoidably contaminated by contributions from the host
galaxy. However, few studies have taken background corrections into
account when calculating the total ring flux, except for
\citet[][]{maoz1996}, who measured the total ring flux by integrating
all counts above the background over the entire ring region. Note that
their background measure is, in fact, a constant value, determined in
an `empty' corner of the image. This thus only provides a simple
approximation to the often highly complex background in the ring
region. Consequently, the main motivation for this paper is to
demonstrate the feasibility of a newly devised method to obtain clean
ring luminosities based on galaxy light-profile fitting using the {\sc
  galfit} algorithm \citep[][]{peng2002,peng2010}.

As a test case, we will obtain the spectral energy distribution (SED)
and SFR of the luminous nuclear ring in the galaxy NGC 1512, while
simultaneously allowing for inherent dust attenuation and background
correction. Our analysis is based on multi-waveband {\sl Hubble Space
  Telescope} ({\sl HST}) imaging data, combined with observations
obtained with the Infrared Array Camera (IRAC) on board the {\sl
  Spitzer} Space Telescope. NGC 1512 is located at high Galactic
latitude and it is, hence, barely affected by Galactic foreground
extinction, $E(B-V)$ = 0.011 mag \citep[][]{schlegel1998}; we will
hence ignore any correction for Galactic foreground reddening. The
remainder of this paper is organized as follows. In Section 2, we
describe the observational data used as well as our data reduction
approach. Section 3 presents in detail our improved procedure for
calculating the ring flux, and addresses a series of technical
issues. Finally, we provide a brief summary and our main conclusions
in Section 4.

\section{Observations and data reduction}

Broad-band {\sl HST}/Wide Field Camera-3 (WFC3; pixel size $\sim 0.04
\arcsec$) images of NGC 1512 were acquired from the {\sl HST} Legacy
Archive (HLA)\footnote{http://hla.stsci.edu/hlaview.html}, observed as
part of program GO-13364 (PI: Calzetti). Our multi-wavelength imaging
data set included observations through the F336W (roughly
corresponding to the Johnson--Cousins $U$ band), F438W ($B$), F555W
($V$), and F814W ($I$) filters, which were pipeline-processed and
calibrated using the standard HLA reduction software. They were
already aligned to the same orientation and field of view, covering
the inner galactic region (see Fig. 1). In this paper we assume a
distance modulus of $(m-M)_0 = 30.48\pm 0.25$ mag\footnote{This value,
  computed from values collected in the NASA Extragalactic Database
  (NED; http://nedwww.ipac.caltech.edu/), presents the geometric mean
  of 10 individual distance measurements from the literature.}, which
corresponds to a distance of 13 Mpc. This results in a pixel scale of
2.4 pc pixel$^{-1}$. Since the circumnuclear starburst ring represents
a small fraction of the full image, for convenience we cropped the
ring-dominated area in the original four exposures to yield a final,
common science image of $585 \times 585$\,pixels$^2$ (or $1400 \times
1400$\,pc$^2$), as shown in Fig. 1.

We used \citet{kennicutt2009}'s calibration of the local SFR:
\begin{equation}
\rm{SFR}\, (M_{\odot}\, yr^{-1}) = 5.5 \times 10^{-42} [{\it L}
  (H\alpha)_{\rm{obs}} + 0.011{\it L}(8\,{\mu}m)],
\end{equation}
where {\it L}(H$\alpha$)$_{\rm{obs}}$ is expressed in units of erg
s$^{-1}$ prior to any internal dust-attenuation correction, and
$L(8\,\mu$m) is the {\sl Spitzer} 8 $\mu$m polycyclic aromatic
hydrocarbon (PAH) emission-line luminosity, also in erg s$^{-1}$. This
calibration was derived assuming a Kroupa initial mass function (IMF)
with stellar masses in the range 0.1--100 $M_{\odot}$
\citep[][]{kroupa2003} and solar chemical abundance;
\citet{allard2006} and \citet{sarzi2007} showed that circumnuclear
regions of barred galaxies can be modeled well by adopting solar
metallicity. H$\alpha$ emission is dominated by young massive stars
and is commonly used as an instantaneous SFR indicator, tracing stars
with lifetimes of $\sim$3--10 Myr and masses greater than 40
$M_{\odot}$ \citep[e.g.,][]{kennicutt2009,hao2011}. Since {\sl HST}
observations of the NGC 1512 circumnuclear ring in H$\alpha$ line
emission are also available in the HLA, we additionally retrieved a
narrow-band F658N (H$\alpha$) image (GO-6738; PI: Filippenko),
observed with the WFPC2/Planetary Camera (PC). The PC chip has a pixel
size of $\sim 0.05 \arcsec$. The pixel scale of the H$\alpha$ image
was resampled to match that of the WFC3 images ($\sim 0.04 \arcsec$)
and then aligned to the F555W frame, using standard {\sc
  iraf/stsdas}\footnote{The Image Redcution and Analysis Facility
  ({\sc iraf}) is distributed by the National Optical Astronomy
  Observatories, which are operated by the Association of Universities
  for Research in Astronomy, Inc., under cooperative agreement with
  the US National Science Foundation.} routines. Observational
information for all {\sl HST} images used in this paper can be found
in Table 1.

The aligned H$\alpha$ band is actually a combined emission line plus
continuum image. To remove the underlying stellar continuum, we used
the scaled F555W and F814W images on both the short- and
long-wavelength sides of the F658N filter bandpass to linearly
interpolate a continuum image at the reference H$\alpha$
wavelength. Since the WFC3 and WFPC2/PC detectors have almost the same
PSF size, FWHM $\sim 0.07 \arcsec$ (i.e., they have comparable spatial
resolutions), we did not apply PSF matching. Next, the continuum image
thus generated was subtracted from the F658N image, after converting
both images from counts s$^{-1}$ to the physical flux units of erg
s$^{-1}$ cm$^{-2}$ \AA$^{-1}$ using the calibration parameter {\sc
  photflam} (i.e., the inverse sensitivity, contained in the image
header), also listed in Table 1. This provides us with a reliable
continuum subtraction without generating any conspicuous negative-flux
regions, as shown in left-bottom panel of Fig. 1.  As a
  consistency check, we compared our continuum-subtracted H$\alpha$
  image to Fig. 1 of \citet[][]{maoz2001}. The latter was generated
  based on their WFPC2/F547M and F814W images. We changed the display
  colors, inverted the associated color map, and adjusted the scale
  limits so as to match the appearance of Maoz et al.'s image as
  closely as possible. Although both images are indeed very similar,
  our new, higher-resolution image shows additional details in the
  form of dust lanes. This will, however, not cause any problems in
  measuring the ring flux, since the faint continuum structure forms
  part of the background flux, which must be subtracted from the image
  (see below). We prefer to use the WFC3/F555W filter rather than
  WFPC2/F547M, because the more recently installed WFC3 camera is
  characterized by better spatial resolution and improved
  efficiency. Moreover, the F555W image reaches fainter photoemtric
  limits and hence it shows more detailed structures compared with
  F547M. Note that our continuum-subtracted H$\alpha$ image includes
adjacent [N{\sc ii}] lines redshifted to $\lambda \lambda 6568,
6604$\AA. We adopted a redshift of 0.0029 \citep[][]{kori2004}.

\begin{figure}
\begin{center}
\includegraphics[width=0.98\columnwidth]{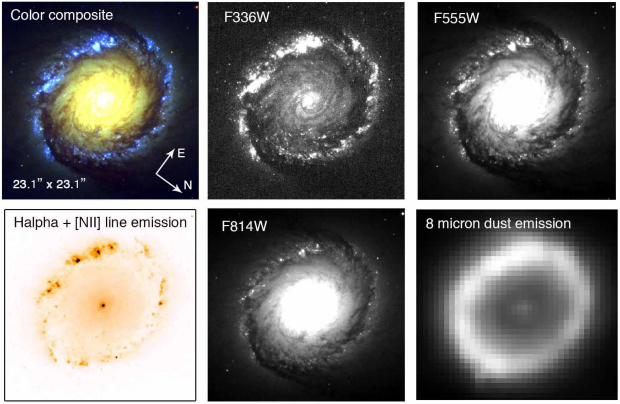}
\caption{Top left: Color composite image of our target $585 \times
  585$\,pixels$^2$ nuclear ring area in NGC 1512, created by stacking
  images taken through the F336W, F555W, and F814W filters. The F336W
  filter, which traces young stellar populations, is shown in blue,
  while F555W and F814W are shown in green and red, respectively. The
  F555W and F814W images are shown in the middle and right-hand
  panels, respectively. Bottom: Continuum-subtracted H$\alpha$/[N{\sc
      ii}] (left), F814W (middle), and continuum-subtracted 8 $\mu$m
  (right) images. All panels share the same orientation and field of
  view as indicated in the top left-hand panel.}
\end{center}
\end{figure}

\begin{table*}
\begin{threeparttable}
\begin{center}
\begin{minipage}{160mm}
\caption{Observational details of the final NGC 1512 image data set.}
\begin{tabular}{|r|lllcll|}
\hline
\multirow{7}{0.8cm}{\sl HST} 
&Filter&Proposal ID/PI&Camera & Exposure time (s) &{\sc photflam}&Zmag\\
\cline{2-7}
&F336W ($U$) & GO-13364/Calzetti & WFC3 &1107& $1.3168305\times10^{-18}$&23.6012 \\
&F438W ($B$) & GO-13364/Calzetti & WFC3  &953& $6.91138715\times10^{-19}$&24.3011\\
&F555W ($V$) & GO-13364/Calzetti & WFC3 &1131& $1.876528\times10^{-19}$ &25.7166\\
&F658N(H$\alpha$) & GO-6738/Filippenko& WFPC2/PC&5200&$1.45442135\times10^{17}$&20.993\\
&F814W($I$) & GO-13364/Calzetti & WFC3 & 977& $1.5304799\times10^{-19}$&25.9379\\
\hline
\multirow{3}{0.8cm}{\sl Spitzer}
&Band&Program ID/PI&Camera & Exposure time (s) &FLUXCONV&Zmag\\
\cline{2-7}
&3.6 $\mu$m & SINGS1-159/Kennicutt & IRAC &193&0.1069&18.8024 \\ 
&8 $\mu$m & SINGS1-159/Kennicutt & IRAC &176&0.2026&17.1984 \\
\hline
\end{tabular}
\begin{tablenotes}
\item {\bf Notes}. Keword {\sc photflam} is given in units of erg
  s$^{-1}$ cm$^{-2}$ \AA$^{-1}$. The instrumental magnitudes have been
  converted to the {\sc STmag} magnitude system by applying the
  photometric zero-points (Zmag), whose calibration relies on the {\sc
    photflam} and {\sc photzpt} keywords, i.e., Zmag = $-$2.5 log({\sc
    photflam}) + {\sc photzpt} (where {\sc photzpt}= $-$21.10 for the
  ST magnitude scale). The {\sl Spitzer} Zmags were taken directly
  from the image headers.
\end{tablenotes}
\end{minipage}
\end{center}
\end{threeparttable}
\end{table*}

However, the possible effects of dust extinction in the H$\alpha$
filter significantly undermine its reliability for measuring the
SFR. Dust particles in the interstellar medium absorb a fraction of
the H$\alpha$ flux; the attenuated H$\alpha$ luminosity is
subsequently re-radiated at infrared wavelengths. To more robustly
estimate the dust-corrected SFRs, a better solution involves combining
H$\alpha$ with infrared observations. We used the {\sl Spitzer}/IRAC
post-basic calibrated data (`post-BCD') images (3.6 and 8 $\mu$m
bands, with a pixel scale of $0.6\arcsec$; see Table 1) taken from the
{\sl Spitzer} Heritage Archive and observed as part of the {\sl
  Spitzer} Infrared Nearby Galaxies Survey
\citep[SINGS;][]{kennicutt2003}. To derive the pure 8 $\mu$m PAH
emission, needed as input in Eq. (2), we had to remove the stellar
continuum contribution from the 8 $\mu$m image. Since the 3.6
  $\mu$m band traces stellar mass in nearby galaxies
\citep[e.g.,][]{elmegreen1984,eskew2012,meidt2012}, a continuum-free
PAH emission image at 8 $\mu$m can be obtained by subtracting a scaled
3.6 $\mu$m band image from the original 8 $\mu$m image, following the
recipe of \citet[][]{helou2004} and employing a scale factor of
0.255. This approach results in highly accurate continuum-subtracted
images, which has been widely demonstrated and used in the literature
\citep[e.g.,][]{calzetti2007,kennicutt2009,zhou2015}. The bottom
right-hand panel of Fig. 1 shows the continuum-subtracted 8 $\mu$m
image cropped to focus on the inner ring area only.

\section{Measuring the integrated ring luminosities}

To characterize the ring structure, we employed {\sc galfit} to
simultaneously fit the two-dimensional stellar light distributions of
multiple galactic components using analytical functions. It generates
a final model image based on the best-fitting structural
parameters. The optimal solution for the model parameters is obtained
using the iterative Levenberg--Marquardt algorithm
\citep[e.g.,][]{bevington2003}. Since the standard statistical
uncertainties returned by {\sc galfit} tend to underestimate the true
uncertainties, we evaluated the goodness-of-fit based on both the
$\chi^2$ values given by {\sc galfit} and visual inspection of the
residual images.

\begin{figure}
\begin{center}
\includegraphics[width=1.0\columnwidth]{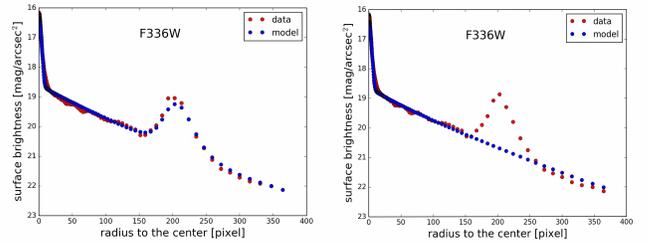}
\caption{ Radial profiles derived from the F336W image. Red dots
  represent the observed data and blue dots are for the best-fitting
  model. Left: Best-fitting model using three galactic components
  (i.e., background and nuclear ring). Right: Application of only a
  model of the background.}
\end{center}
\end{figure}

\subsection{{\sl HST} image analysis}

With regard to optical {\sl HST} images, we have developed an
  improved approach to derive the total ring luminosity, based on
inspection of the radial surface brightness profiles. To start with,
we performed a test using the F336W image, which contains the most
luminous nuclear ring component among our optical {\sl HST} bands.

We plotted the radial surface brightness profile in the F336W image,
shown as the red dotted line in the left-hand panel of Fig. 2. The
presence of the nuclear ring introduces a prominent peak at radii
between 118 and 250 pixels, corresponding to a radial range from
4.6$\arcsec$ to 9.8$\arcsec$. Therefore, we adopted this annulus to
define the ring region in this paper. Based on the shape of the
observed radial profile, we assume that our observed image consists of
the host galaxy's background contribution and a nuclear ring, and we
intend to model all galactic components in the original image. The
nuclear ring can be quantified by a truncation function, and the
background contribution in the image can be fitted by adopting a
mixture of Gaussian and S\'{e}rsic functions. This approach yields a
good fit with minimal residuals. The truncation function is, in
essence, a hyperbolic tangent function. In {\sc galfit} these
functions are commonly used to modify the light profiles of galactic
components, such as S\'{e}rsic profiles, to produce ring-like
structures \citep[for a detailed description, see][]{peng2010}.

To limit the degrees of freedom in the fits, we (only) fixed the
center position; the other parameter were left as free parameters. We
found a good match between the best-fitting model and the real
data. The middle panel of Fig. 3 presents our best-fitting model based
on the improved method, and the relevant radial profile is shown
in the left-hand panel of Fig. 2 (blue dotted line; the red line
represents the actual data). It is clear that a three-component model
could entirely recover the observational profile with high
accuracy. Using this method, the total ring luminosity can be derived
by adding the flux in all pixels of the predefined ring region: see
the right-hand panel of Fig. 3, which shows the truncation component
(designed to model the nuclear ring). We determined $m_{\rm{ring}}(U)
= 14.83^{+0.05}_{-0.05}$ mag.

We also investigated how the use of different PSF images affected our
fit results. We ran {\sc galfit} using an empirical PSF, a theoretical
PSF, and without any input PSF, respectively. We verified that the
effect of changing the PSF is negligible, because the three conditions
employed accurately reproduced the model image with the same
structural parameters and the same resulting ring luminosity. This can
be readily understood given the significant difference between the
ring and the PSF sizes: the typical WFC3/UVIS stellar FWHM of $\sim
0.08\arcsec$\citep[e.g.,][]{calzetti2015} is much smaller than the
ring radius $\sim8\arcsec$. To find the best fits faster, we opted to
not provide the PSF when dealing with our {\sl HST} images.

\begin{figure}
\begin{center}
\includegraphics[width=1.0\columnwidth]{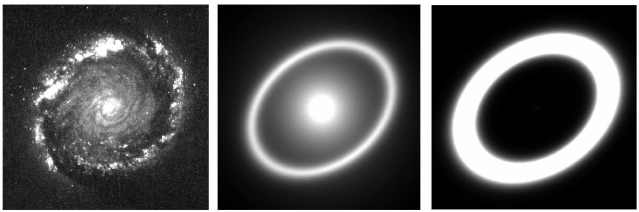}
\caption{Left: Observed F336W image. Middle: Best-fitting
  three-component model for the F336W image. Right: Image of the
  individual truncation component based on the best-fitting model in
  the middle panel.}
\end{center}
\end{figure}

We subsequently applied our method to the other {\sl HST} filters.
The measured total ring magnitudes in the other filters in our data
set are listed in Table 2. Note that no attempt was made to obtain a
definitive model for the bulge. Our main goal was just to measure the
proper, localized ring structure, with the remaining parts of the
image together considered the image's (i.e., the ring's) background.

To determine the observed ring's H$\alpha$ luminosity, we applied our
method to the continuum-subtracted H$\alpha$+[N{\sc ii}] image
produced in the previous section. We found $L({\rm H}\alpha+[{\rm
    NII}])=$ $5.61\times10^{39}$ erg s$^{-1}$, with a fractional
uncertainty of $\sim$10\%. This is very close to the luminosity
(6$\times 10^{39}$ erg s$^{-1}$) derived by \citet[][]{maoz2001},
  although those authors did not actually define a clear-cut ring
  region. Next, we corrected for the contributions of the [N{\sc ii}]
lines using the theoretical [N{\sc ii}] flux line ratio
\citep[1/3;][]{osterbrock2006} and the observed [N{\sc
    ii}]$\lambda$6584/H$\alpha$ flux line ratio of NGC 1512 obtained
from \citet[][]{calzetti2007}. $L({\rm [NII]})$ was removed as a
function of transmission efficiency of the F658N filter at the
position of the lines. We obtained $L_{\rm obs}({\rm H}\alpha)=4.74
\times 10^{39}$ erg s$^{-1}$, which will be used to calculate the
ring's SFR, together with $L(8\,\mu {\rm m})$, discussed in the next
subsection.

\subsection{{\sl Spitzer} image analysis}

\begin{figure}
\begin{center}
\includegraphics[width=1.0\columnwidth]{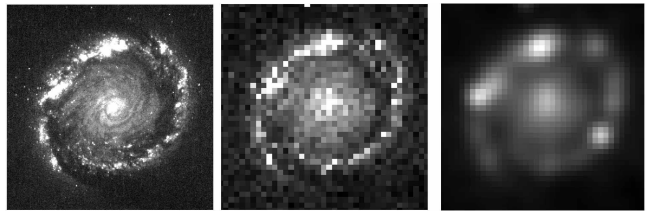}
\caption{Left: Original F336W image. Middle: As the left-hand image,
  but for a pixel scale that has been degraded to that of the infrared
  image. Right: Final image after degradation of its pixel scale and
  Gaussian smoothing, so as to mimic the observed infrared data.}
\end{center}
\end{figure}

A caveat regarding the {\sl Spitzer}/IRAC image is that the IRAC image
resolution for a typical PSF FWHM $\sim 2 \arcsec$ is much lower than
that of the {\sl HST} PSF. As illustrated in the bottom right-hand
panel of Fig. 1, such a resolution is comparable to the radius of the
ring, and hence we cannot ascertain whether or not the measured ring
luminosity would be affected by the lower {\sl Spitzer} resolution. To
clarify this issue, our strategy consisted of artificially degrading
the {\sl HST}/WFC3 F336W image to the resolution of {\sl Spitzer}/IRAC
so as to simulate the conditions pertaining to low resolution. We
subsequently ran {\sc galfit} on the degraded F336W image to verify
whether we could recover the same result as that derived before image
degradation. To construct the simulated image, we first changed its
pixel size to that of {\sl Spitzer} ($\sim 0.6\arcsec$/pixel), as
shown in the middle panel of Fig. 4. It was then broadened by
convolution with a Gaussian kernel to match the smoothness of the {\sl
  Spitzer} image, as shown in the right-hand panel of Fig. 4. We
  found that the ring magnitude derived from our simulated image is
  $m_{\rm{ring}}(U) = 14.83_{-0.17}^{+0.19}$ mag, which is in very
  good agreement with that obtained prior to image degradation,
  supporting the application of our method to {\sl Spitzer} images.

\begin{figure}
\begin{center}
\includegraphics[width=0.9\columnwidth]{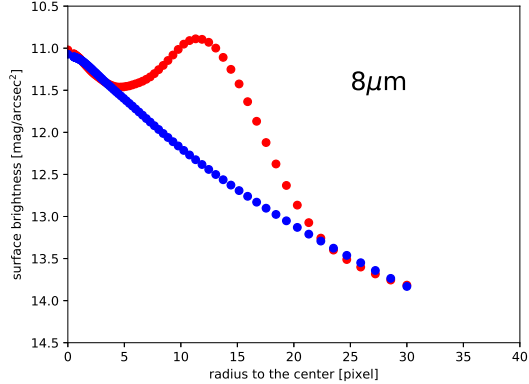}
\caption{Radial profile of the ring in the {\sl Spitzer} 8 $\mu$m
  image. The red dotted line represents the actual data, while the
  blue line is the best-fitting background.}
\end{center}
\end{figure}

We next applied our method to the continuum-subtracted {\sl Spitzer} 8
$\mu$m image, with the input PSF image artificially created by the
{\sc STinyTim} package \citep[][which is a new version of {\sc
    TinyTim} developed specifically for {\sl Spitzer}
  data]{krist2005}. It is important to note that the ring is
morphologically rather extended in the infrared compared with its
extent in the optical images, but we chose the ring region according
to that defined in the previous subsection to obtain its PAH
emission-line luminosity, an essential parameter to determine the
ring's SFR. In essence, our standard ring region was defined on the
basis of the {\sl HST}/F336W image, which traces the young stellar
population. The equation we used to determine the SFR is also
sensitive to the age range (roughly from 3 Myr to 10 Myr).

The resulting luminosity is $L(8\,\mu$m) = $(8.857\pm 0.6) \times
10^{41}$ erg s$^{-1}$. Substituting $L(8\,\mu$m) and $L({\rm
  H\alpha})_{\rm obs}$ into Eq. (2), we derive an extinction-corrected
$L({\rm H\alpha})_{\rm corr}= (1.445\pm 0.17)\times10^{40}$ erg
s$^{-1}$, which is within the estimated range of H$\alpha$+[N{\sc ii}]
luminosity given by \citet[][]{maoz2001}, who provided an
order-of-magnitude estimation in the range from $\sim
10^{40}$--$10^{41}$ erg s$^{-1}$. This indicates a total SFR of 0.08
$M_{\odot}$ yr$^{-1}$ based on Eq. (2). The area of the ring is 0.879
kpc$^2$, so the SFR surface density (i.e., the star formation per unit
area) is $\Sigma_{\rm SFR}= 0.09\, M_{\odot} \,{\rm yr}^{-1}$ ${\rm
  kpc}^{-2}$, which is about an order of magnitude higher than the
disk-averaged SFR densities in normal galaxies
\citep[e.g.,][]{kennicutt1998}.

 For comparison, we also explored the performance of a simpler
  method to determine the ring flux. Here, we tried to obtain the
  best-fitting ring background profile. As before, the background can
  be modeled by S\'{e}rsic and Gaussian functions. However, the ring
  region was masked to prevent overestimation of the background before
  running the fit. The blue dotted line in the right-hand panel of
  Fig. 2 shows the best-fitting background in the F336W data, which
  matches the real background nicely. The total ring luminosity can be
  derived by subtracting the model from the data across the ring
  region, yielding $m_{\rm{ring}}(U)=14.76^{+0.13}_{-0.15}$ mag. The
  associated uncertainties were estimated by considering both the
  fitting errors and the influence of different masking annulus radii
  applied to the ring. This is, within the uncertainties, consistent
  with the value resulting from our more elaborate method: see Table 2
  for comparison.  

We note that the integrated ring luminosity returned by our improved
method (see Section 3.1) is derived completely based on adequate model
assumptions, including the use of a truncation function to model the
ring, but for this simpler method the luminosity is directly derived
from the `real' image. The fact that our two different approaches
reach the same conclusion for this particular image featuring a very
bright ring component demonstrates the feasibility of our more
elaborate method and the robustness of the results under such
circumstances. The ring magnitude is a function of wavelength and age,
however. The nuclear ring gradually becomes less prominent toward
redder wavelengths, and it is almost undetectable in the radial
profile of the F814W image, as shown in Fig. 1. Application of our
simple method would hence lead to unreasonably large uncertainties. It
is therefore only suitable for blue filters and/or very luminous
rings.

Because of the remarkably luminous ring seen in our
continuum-subtracted 8 $\mu$m, H$\alpha$+[N{\sc ii}], and 8 $\mu$m
images, as evidenced in both Fig. 1 and Fig. 5, we proceeded to fit
their backgrounds based on application of the simple model as well
(see Fig. 5 for an example). We found that the resulting flux values
and the SFR derived were consistent with those found using the more
elaborate method: see Table 2. Table 2 also includes the best-fitting
ring parameters (see below).

\begin{table*}
\begin{threeparttable}
\begin{center}
\begin{minipage}{160mm}
\caption{Photometry and physical properties of the NGC 1512 ring}
\label{comparison.tab}
\begin{tabular}{ccccccc}
\hline
\hline \\[-1.3ex]
\colhead{H$\alpha$+[N{\sc ii}]}&\colhead{F336W}&\colhead{F438W}&\colhead{F555W}&\colhead{F814W}&\colhead{8$\mu$m}&continuum-subtracted 8 $\mu$m\\[0.5ex]

\hline\\[-1.4ex]
$13.93 \pm 0.06$&$14.83 \pm 0.05$&$15.23 \pm 0.08$&$16.12^{+0.12}_{-0.10}$&$17.14 \pm 0.17$&$5.58 \pm 0.27$&$5.60 \pm 0.28$\\[1.2ex]
$13.87^{+0.10}_{-0.09}$&$14.76^{+0.13}_{-0.15}$&---&---&---&$5.66 \pm 0.31$&$5.69 \pm 0.31$\\[1.2ex]

\hline
\hline\\[-1.3ex] 
&log ($t$ yr$^{-1}$)&log (${M}/{M}_{\odot}$)&{\tt dust1}&{\tt dust2}&\\[0.6ex]
\hline\\[-1.5ex]
Best-fitting parameters &7.63$^{+0.15}_{-0.31}$&7.1 $\pm$ 0.11&1.06$^{+0.68}_{-0.65}$&0.57$_{-0.20}^{+0.22}$&\\[1.2ex]
\hline
\hline\\[-1ex]
($M/L$)$\times L$&7.17&7.15&7.08&7.2&7.72\\[0.5ex]
\hline
\end{tabular}
\begin{tablenotes}
\item {\bf Notes}. We do not provide the photometric results from the
  simple method for redder optical filters, because of the
  intrinsically large uncertainties (see the text). The logarithmic
  mass estimates, based on the $M/L$s calculated, are also given as a
  function of filter/wavelength. {\tt dust1} and {\tt dust2} are
  dimensionless parameters.
\end{tablenotes}

\end{minipage}
\end{center}
\end{threeparttable}
\end{table*}

\subsection{Spectral energy distribution modeling}

\begin{figure}
\begin{center}
\includegraphics[width=1.0\columnwidth]{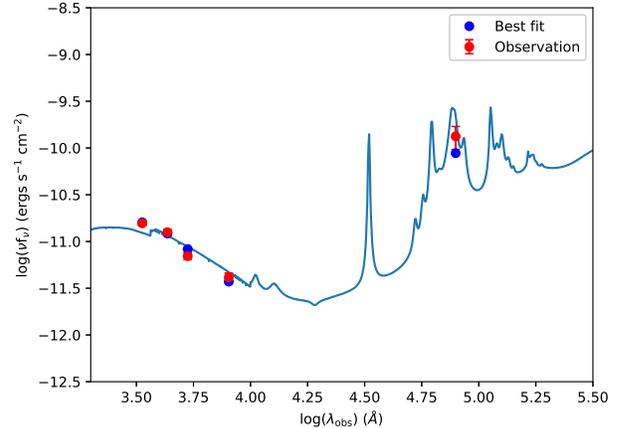}
\caption{Spectral energy distribution. The observed broadband ring
  fluxes are shown as red dots, along with the photometry (blue dots)
  and the best-fitting model spectrum (blue line). Because of the
  relatively low resolution of {\sl Spitzer}, the fit in the 8 $\mu$m
  band has a larger uncertainty.}
\end{center}
\end{figure}

We next modeled the observed SED using Flexible Stellar Population
Synthesis models \citep[FSPS;][]{conroy2009,conroy2010} to derive the
ring's average age and total stellar mass. We did not correct for
emission-line contamination to the broad-band magnitudes, because this
minor effect can be ignored. The stellar population synthesis model
templates were internally generated over a grid of stellar population
parameters, with the metallicity again fixed to the solar value. We
assumed a Kroupa (2001) IMF, and left the age, mass, and dust
parameters free in the fits. In FSPS models, the (dimensionless) dust
parameters {\tt dust1} and {\tt dust2} describe the attenuation of
young and old stellar light, respectively \citep[for a detailed
  description, see][]{conroy2009}, corresponding to ${\hat \tau_1}$
and ${\hat \tau_2}$ in \citet[][]{charlot2000}'s prescription (their
Eq. 5.8). The priors of our free parameters were 6.0 $<$ log
(${M}/{M}_{\odot}$) $<$ 9.0, 6.0 $<$ log ($t$/yr) $<$ 10, 0.1 $<$ {\tt
  dust1} $<$ 2.0, and 0.1 $<$ {\tt dust2} $<$ 2.0. For each parameter,
a posterior probability distribution was constructed by employing
Markov chain Monte Carlo methodology.

Figure 6 shows our best-fitting result; the photometry is represented
by the blue dots, while the solid line is the model spectrum based on
the best-fitting parameters. The best-fitting stellar mass is log
(${M}/{M}_{\odot}$) = 7.1 $\pm$ 0.11, and the error bars are the
16$^{\rm th}$ and 84$^{\rm th}$ percentiles of the model
posteriors. This mass value corresponds to approximately 0.004\% and
0.2\% of the total dynamical and the H{\sc i} masses of NGC 1512
\citep[e.g.,][]{kori2009}, respectively. To further examine the
reliability of the derived mass, we calculated the mass-to-light ratio
($M/L$) based on the best-fitting parameters. The mass can then be
estimated approximately by multiplying by our observed photometry. As
shown in Table 2, we found that the derived masses based on the
best-fitting $M/L$ ratios are consistent within the 1$\sigma$ error
ranges in our four {\sl HST} filters, with the 8 $\mu$m band being the
only outlier. This is mainly due to the larger observational
uncertainty in that filter. This reflects the robustness of our mass
determination. The best-fitting average age is log ($t$ yr$^{-1}$) =
7.63$^{+0.15}_{-0.31}$, in agreement with the young-starburst
characteristics of the ring. Given the present-day SFR derived above
and the ring's average age, the total stellar mass formed over its 40
Myr lifetime assuming a constant SFR, would amount to
$\sim$$3.8\times10^6 {M}_{\odot}$. This only accounts for
approximately one-quarter of the best-fitting total ring mass. This
insight thus demonstrates that the nuclear ring's star-formation
history is likely more complex. There is some evidence that the
observed emission lines in the ring might be best modeled by adopting
multiple starburst episodes of varying intensity rather than by a
constant SFR \citep[e.g.,][]{allard2006,sarzi2007}.

\section{Summary and conclusions}

Based on the publicly available optical {\sl HST}/WFC3 and infrared
{\sl Spitzer}/IRAC images, we have presented {an improved} method
(based on the {\sc galfit} code) to derive the SED and SFR of the
conspicuous nuclear ring in the barred spiral galaxy NGC 1512. The
good agreement between the results from our improved method with
  that from a simple two-component method indicates that it is
possible to accurately measure the integrated ring properties. We
derived the SFR following the prescription given by
\citet[][]{kennicutt2009}. Based on Eq. (2), this offers a composite
(multi-wavelength) H$\alpha$+8 $\mu$m-based SFR which is
attenuation-corrected (composite methods provide more robust SFRs than
any of the single-wavelength methods). We treated the contribution of
the background intrinsic to the host galaxy more carefully, through
modeling the light profile in the observed images. These two
  aspects represent the most important novel aspects of our study,
  since they enable us to obtain more realistic SFRs characterized by
  smaller uncertainties than those derived in previous studies.
 
All nuclear components of the galaxy in the image were modeled using
analytical functions. To derive the ring's current SFR, we constructed
high-quality continuum-subtracted H$\alpha$ and 8 $\mu$m images. The
resulting SFR surface density ($\Sigma_{\rm SFR}$) is $0.09\,
{M}_{\odot} \,{\rm yr}^{-1}$ ${\rm kpc}^{-2}$. Our approach represents
a significant improvement in measurement accuracy compared with
previous efforts in the literature, which are often based on simple
assumptions. Finally, we compared the observed ring's SED with FSPS
model SEDs to derive its physical properties, including its average
age and total stellar mass. The NGC 1512 ring has a total stellar mass
of $\sim$ $10^7 {M}_{\odot}$ and a young average age of around 40 Myr.

We have compiled a catalog of nuclear rings which have already been
observed and are well resolved by both the {\sl HST} (in at least four
filters, ideal for further star cluster analysis) and {\sl Spitzer}
telescopes. In an extensive follow-up study, we are working on
application of our improved method to a statistically carefully
selected sample of nuclear rings. We will combine these results with
the detected cluster populations to explore possible relationships
between ring star-formation properties and those of the young cluster
populations (e.g., cluster luminosity and mass functions, the cluster
mass fraction, and cluster formation efficiencies). Our main objective
in this paper was, therefore, to develop essential technical tools for
our follow-up statistical study.

\section*{Acknowledgments}

We thank Hua Gao and Jinyi Shangguan for their generous help with
technical issues and valuable discussions. This paper is based on
observations made with the NASA/ESA {\sl Hubble Space Telescope} and
obtained from the {\sl Hubble} Legacy Archive, which is a
collaboration between the Space Telescope Science Institute
(STScI/NASA), the Space Telescope European Coordinating Facility
(ST-ECF/ESA), and the Canadian Astronomy Data Centre
(CADC/NRC/CSA). This work also uses observations made with the {\sl
  Spitzer Space Telescope}, which is operated by the Jet Propulsion
Laboratory, California Institute of Technology, under a contract with
NASA. This research has also made use of NASA's Astrophysics Data
System Abstract Service. We acknowledge research support from the
National Natural Science Foundation of China through grants U1631102,
11373010, and 11633005 (C. M. and R. d. G.) and 11473002 and 11303008
(L. C. H.). L. C. H. also acknowledges support from the National Key
Program for Science and Technology Research and Development, grant
2016YFA0400702.


\begin{thebibliography}{}
\bibitem[Athanassoula(1992)]{Athan1992} Athanassoula, E. 1992, MNRAS,
  259, 345
\bibitem[Athanassoula(1994)]{Athan1994}Athanassoula, E., 1994, in
  Shlosman I., ed., Mass-Transfer Induced Activity in
  Galaxies. Cambridge Univ. Press, Cambridge, p. 143
\bibitem[Allard et al.(2006)]{allard2006}Allard, E. L., Knapen, J. H.,
  Peletier, R. F., \& Sarzi, M. 2006, MNRAS, 371, 1087
\bibitem[()]{}Anders P., Fritze-v. Alvensleben U., 2003, A\&A, 401, 1063 
\bibitem[Barth et al.(1995)]{barth1995} Barth, A. J., Ho, L. C.,
  Filippenko, A. V., \& Sargent, W. L. W. 1995, AJ, 110, 1009
\bibitem[Benedict et al.(2002)]{benedict2002}Benedict, G. F., Howell,
  D. A., J\o rgensen, I., Kenney, J. D. P., \& Smith, B. J. 2002, \aj,
  123, 1411
\bibitem[Bevington \& Robinson(2003)]{bevington2003}Bevington, P. R.,
  \& Robinson, D. K. 2003, Data Reduction and Error Analysis for the
  Physical Sciences (3rd ed.; Boston: McGraw-Hill)
\bibitem[Burbidge \& Burbidge(1960)]{burbidge1960}Burbidge, E. M., \&
  Burbidge, G. R. 1960, \apj, 132, 30
\bibitem[Buta \& Combes(1996)]{Buta1996}Buta, R., \& Combes, F. 1996,
  FCPh, 17, 95
\bibitem[Buta \& Crocker(1993)]{buta1993}Buta, R., \& Crocker,
  D. A. 1993, \aj, 105, 1344
\bibitem[Buta et al.(1999)]{buta1999}Buta, R., Crocker D. A., Byrd,
  G. G., 1999, AJ, 118, 2071
\bibitem[Buta et al.(2000)]{buta2000}Buta, R., Treuthardt, P. M.,
  Byrd, G. G., \& Crocker, D. A. 2000, \aj, 120, 1289
\bibitem[Byrd et al.(1994)]{Byrd1994} Byrd, G., Rautianen, P., Salo,
  H., Buta, R., \& Crocker, D.A. 1994, \aj, 108, 476
\bibitem[Calzetti et al.(2007)]{calzetti2007}Calzetti D. et al., 2007,
  ApJ, 666, 870
\bibitem[Calzetti et al.(2015)]{calzetti2015}Calzetti, D., Lee, J. C.,
  Sabbi, E., et al., 2015, AJ, 149, 51
\bibitem[Charlot \& Fall(2000)]{charlot2000}Charlot S., \& Fall,
  S. M., 2000, ApJ, 539, 718
\bibitem[Comer\'{o}n et al.(2010)]{comeron2010}Comer\'{o}n, S.,
  Knapen, J. H., Beckman, J. E., et al. 2010, MNRAS, 402, 2462
\bibitem[Combes(2001)]{combes2001}Combes, F., 2001, in Aretxaga I.,
  Kunth, D., M\'{u}jica, R., eds, Advanced Lectures on the
  Starburst--AGN Connection. World Scientific, Singapore, p. 223
\bibitem[Conroy \& Gunn(2010)]{conroy2010}Conroy, C., \& Gunn, J. E.,
  2010, ApJ, 712, 833
\bibitem[Conroy et al.(2009)]{conroy2009}Conroy, C., Gunn, J. E.,
  White, M., 2009, ApJ, 699, 486
\bibitem[de Grijs \& Anders (2012)]{grijs2012}de Grijs, R., \& Anders,
  P., 2012, ApJ, 758, L22
\bibitem[de Grijs et al.(2017)]{grijs2017}de Grijs, R., Ma, C., Jia,
  S. Y., Ho, L. C., \& Anders, P., 2017, MNRAS, 465, 2820
\bibitem[Elmegreen \& Elmegreen(1984)]{elmegreen1984}Elmegreen D. M.,
  \& Elmegreen B. G., 1984, ApJS, 54, 127
\bibitem[Englmaier \& Shlosman(2000)]{englmaier2000}Englmaier, P., \&
  Shlosman, I. 2000, \apj, 528, 677
\bibitem[Eskew et al.(2012)]{eskew2012}Eskew, M., Zaritsky, D., \&
  Meidt, S., 2012, AJ, 143, 139
\bibitem[Genzel et al.(1995)]{genzel1995}Genzel, R., Weitzel, L.,
  Tacconi-Garman, L. E., Blietz, M., Cameron, M., Krabbe, A., Lutz,
  D., \& Sternberg, A. 1995, \apj, 444, 129
\bibitem[Hao et al.(2011)]{hao2011}Hao, C.-N., Kennicutt, R. C.,
  Johnson, B. D., et al. 2011, ApJ, 741, 124
\bibitem[Helou et al.(2004)]{helou2004}Helou, G., et al. 2004, ApJS,
  154, 253
\bibitem[Hsieh et al.(2012)]{hsieh2012}Hsieh, P.-Y., Ho, P. T. P.,
  Kohno, K., Hwang, C.-Y., \& Matsushita, S. 2012, \apj, 747, 90
\bibitem[Hummel et al.(1987)]{hummel1987}Hummel, E., van der Hulst,
  J.M., \& Keel, W. C. 1987, \aap, 172, 32
\bibitem[Kennicutt(1983)]{kennicutt1983}Kennicutt, R. C., Jr., 1983,
  ApJ, 272, 54
\bibitem[Kennicutt(1998)]{kennicutt1998}Kennicutt R. C., Jr, 1998,
  ApJ, 498, 541
\bibitem[Kennicutt(1998)]{kennicutt1998}Kennicutt, R. C. 1998, ARA\&A,
  36, 189
\bibitem[Kennicutt et al.(2003)]{kennicutt2003}Kennicutt, R. C., Jr et
  al., 2003, PASP, 115, 928
\bibitem[Kennicutt et al.(2009)]{kennicutt2009}Kennicutt, R. C., Hao,
  C.-N., Calzetti, D., et al. 2009, ApJ, 703, 1672
\bibitem[Kennicutt \& Evans(2012)]{kennicutt2012} Kennicutt R. C., \&
  Evans N. J., 2012, ARA\&A, 50, 531
\bibitem[Kim et al.(2012a)]{Kim2012a}Kim, W.-T., Seo, W.-Y., \& Kim,
  Y.\ 2012a, \apj, 758, 14
\bibitem[Kim et al.(2012b)]{Kim2012b}Kim, W.-T., Seo, W.-Y., Stone,
  J. M., Yoon, D., \& Teuben, P. J.\ 2012b, \apj, 747, 60
\bibitem[Knapen et al.(1995)]{Knapen1995}Knapen, J. H., Beckman,
  J. E., Heller, C. H., Shlosman, I., \& de Jong, R. S. 1995, \apj,
  454, 623
\bibitem[Knapen et al.(2006)]{Knapen2006}Knapen, J. H., Mazzuca,
  L. M., B\"{o}ker, T., et al. 2006, A\&A, 448, 489
\bibitem[Koribalski \& L$\acute{\rm{o}}$pez-S$\acute{\rm
    a}$nchez(2009)]{kori2009} Koribalski B. S., L$\acute{ \rm
  o}$pez-S$\acute{\rm a}$nchez A. R., 2009, MNRAS, 400, 1749
\bibitem[Koribalski et al.(2004)]{kori2004}Koribalski, B. S.,
  Staveley-Smith, L., Kilborn, V. A., et al. 2004, AJ, 128, 16
\bibitem[Kormendy \& Kennicutt(2004)]{kormendy2004}Kormendy, J., \&
  Kennicutt, R. C., Jr. 2004, ARA\&A, 42, 603
\bibitem[Kroupa \& Weidner(2003)]{kroupa2003}Kroupa, P., \& Weidner,
  C. 2003, ApJ, 598, 1076
\bibitem[Krist et al.(2005)]{krist2005}Krist, J.E., Ardila, D.R.,
  Golimowski, D.A., et al., 2005, AJ, 129, 1008
\bibitem[Li et al.(2015)]{Li2015} Li Z., Shen J., Kim W.-T., 2015,
  ApJ, 806, 150
\bibitem[Maciejewski(2004)]{Mac2004}Maciejewski, W. 2004, \mnras, 354,
  892
\bibitem[Maoz et al.(1996)]{maoz1996} Maoz, D., Barth, A. J.,
  Sternberg, A., et al. 1996, \aj, 111, 2248
\bibitem[Maoz et al.(2001)]{maoz2001}Maoz, D., Barth, A. J., Ho,
  L. C., Sternberg, A., \& Filippenko, A. V. 2001, \aj, 121, 3048
\bibitem[Mazzuca et al.(2008)]{mazzuca2008}Mazzuca, L. M., Knapen,
  J. H., Veilleux, S., \& Regan, M. W. 2008, ApJ, 174, 337
\bibitem[Mazzuca et al.(2006)]{mazzuca2006} Mazzuca, L. M., Sarzi M.,
  Knapen, J. H., Veilleux, S., Swaters, R., 2006, ApJ, 649, 79
\bibitem[Mazzuca et al.(2011)]{mazzuca2011}Mazzuca, L. M., Swaters,
  R. A., Knapen, J. H., \& Veilleux, S. 2011, ApJ, 739, 104
  
\bibitem[Meidt et al.(2012)]{meidt2012}Meidt S. E. et al., 2012, ApJ, 744, 17  
  
\bibitem[Osterbrock \& Ferland (2006)]{osterbrock2006}Osterbrock
  D. E., Ferland G. J., 2006, Astrophysics of Gaseous Nebulae and
  Active Galactic Nuclei, 2nd edn. University Science Books,
  Sausalito, CA
\bibitem[Peng et al.(2002)]{peng2002}Peng, C. Y., Ho, L. C., Impey,
  C. D., \& Rix, H.-W. 2002, AJ, 124, 266
\bibitem[Peng et al.(2010)]{peng2010}Peng, C. Y., Ho, L. C., Impey,
  C. D., et al. 2010, AJ, 139, 2097
\bibitem[Regan \& Teuben(2003)]{Regan2003}Regan, M. W., \& Teuben,
  P. 2003, \apj, 582, 723
\bibitem[Sandage(1961)]{sandage1961}Sandage, A. R. 1961, The Hubble
  Atlas of Galaxies (Washington, DC: Carnegie Institution of
  Washington)
\bibitem[Sarzi et al.(2007)]{sarzi2007} Sarzi, M., Allard, E. L,
  Knapen, J. H., \& Mazzuca, L. M. 2007, MNRAS, 380, 949
\bibitem[Schlegel et al.(1998)]{schlegel1998}Schlegel, D. J.,
  Finkbeiner, D. P., \& Davis, M. 1998, ApJ, 500, 525
\bibitem[Shlosman et al.(1990)]{shlosman1990} Shlosman, I., Begelman,
  M. C., Frank J., 1990, Natur, 345, 679
\bibitem[()]{}S\'{e}rsic J. L., 1968, Atlas de galaxias australes
\bibitem[Sil'chenko \& Moiseev (2006) ]{silchenko2006}Sil'chenko,
  O. K., \& Moiseev, A. V., 2006, AJ, 131, 1336
\bibitem[van der Laan et al.(2013)]{van2013} van der Laan, T. P. R.,
  Schinnerer, E., Emsellem, E., et al. 2013, \aap, 551, A81
\bibitem[Weiner et al.(2001)]{weiner2001}Weiner, B. J., Sellwood,
  J. A., \& Williams, T. B. 2001, ApJ, 546, 931
\bibitem[Wild et al.(1992)]{wild1992}Wild, W., Harris, A. I., Eckart,
  A., Genzel, R., Graf, U. U., et al. 1992, \aap, 265, 44

\bibitem[Willick et al.(1997)]{willick1997}Willick, J. A., Courteau, S., Faber, S. M., et al. 1997, ApJS, 109, 333  
\bibitem[Zhou et al.(2015)]{zhou2015} Zhou, Z.-M., Cao, C., \& Wu,
  H. 2015, AJ, 149, 1
\end{thebibliography}
\end{document}